\documentclass[10pt,conference]{IEEEtran}
\usepackage{latexsym,amssymb,amsmath,graphicx,bbm}
\usepackage{epic}
\usepackage{psfrag}
\usepackage{amsbsy}
\usepackage{mathptm} 
\usepackage[bbgreekl]{mathbbol}
\usepackage{epsfig}
\usepackage{cite}
\usepackage[small,normal,bf,up]{caption2}
\usepackage{setspace}
\usepackage{color}

\newcommand{\brc}[1]{\left({#1}\right)}
\newcommand{\sqbrc}[1]{\left[{#1}\right]}

\newcommand{\bh}{B}
\newcommand{\gbsc}{\mathrm{gbsc}}
\newcommand{\dens}[1]{\mathsf{#1}}
\newcommand{\Ldens}[1]{\dens{#1}}
\newcommand{\ld}[1]{\dens{#1}}
\newcommand{\conv}{\circledast}
\newtheorem{lemma}{Lemma}
\DeclareMathOperator{\prob}{P}
\newcommand{\perr}{\prob_{\text{\rm e}}}
\begin{document}
\title{Degree Optimization and Stability Condition for the Min-Sum Decoder}
\author{
\authorblockN{Kapil Bhattad}
\authorblockA{ECE Department \\
Texas A\&M University\\
College Station, TX 77843\\
kbhattad@ece.tamu.edu}
\and
\authorblockN{Vishwambhar Rathi}
\authorblockA{School of Computer and \\
Communication Sciences, EPFL\\
Email: vishwambhar.rathi@epfl.ch}
\and
\authorblockN{Ruediger Urbanke}
\authorblockA{School of Computer and \\ 
Communication Sciences, EPFL\\
Email: ruediger.urbanke@epfl.ch}
}

\maketitle
\begin{abstract}
The min-sum (MS) algorithm is arguably the second most fundamental algorithm in the
realm of message passing due to its optimality (for a tree code) with respect
to the {\em block error} probability \cite{Wiberg}. There also seems to be a
fundamental relationship of MS decoding with the linear programming decoder
\cite{Koetter}. Despite its importance, its fundamental properties have not
nearly been studied as well as those of the sum-product (also known as BP)
algorithm.

We address two questions related to the MS rule. First, we characterize the
stability condition under MS decoding. It turns out to be essentially the same
condition as under BP decoding. Second, we perform a degree distribution
optimization. Contrary to the case of BP decoding, under MS decoding the
thresholds of the best degree distributions for standard irregular LDPC
ensembles are significantly bounded away from the Shannon threshold.  More
precisely, on the AWGN channel, for the best codes that we find, the gap to
capacity is $1$dB for a rate $0.3$ code and it is $0.4$dB when the rate is
$0.9$ (the gap decreases monotonically as we increase the rate). 

We also used the optimization procedure to design codes for modified MS
algorithm where the output of the check node is scaled by a constant
$1/\alpha$. For $\alpha = 1.25$, we observed that the gap to capacity was
lesser for the modified MS algorithm when compared with the MS algorithm.
However, it was still quite large, varying from 0.75 dB to 0.2 dB for rates
between 0.3 and 0.9.

We conclude by posing what we consider to be the most important open questions
related to the MS algorithm.
\end{abstract}

\section{Introduction}
The min-sum (MS) decoder is perhaps the second most fundamental message passing
decoder after Belief Propagation (BP) decoder for two main reasons. Firstly,
the MS decoder is optimal with respect to block error probability on a tree
code \cite{Wiberg}.  Secondly, it is widely believed that the MS decoder is
closely related to the linear programming (LP) based decoder proposed in
\cite{Feldman}. In \cite{Koetter}, a complete characterization of the decoding
region of the LP decoder has been provided with respect to the pseudocodewords
of the underlying bipartite graph. The results in \cite{Vontobel} suggest that
the decoding region of the LP decoder is identical to that of the MS decoder
(indeed, this is the case for tree codes).  In addition, the MS decoder is of
practical interest because of its low implementation complexity. 

In \cite{Anastasopoulos}, the asymptotic performance of the MS decoder using
density evolution was evaluated. Not much is known, however, analytically about
the density evolution behavior of the MS decoder as compared to BP.  

We first address the issue of {\em stability} of the MS decoder. In particular,
we derive a condition which guarantees that the densities corresponding to the
MS decoder which one observes in density evolution converge to an
``error-free'' density.  This condition turns out to be essentially the same as
the stability condition for BP.  

Recall that for the BP decoder the space of densities which arise in the
context of density evolution is the space of {\em symmetric} densities.  Under
MS decoding, on the contrary, no equivalent condition is known.  Empirically,
one observes that for $y \geq 0$ the densities fulfill the inequality
\begin{align*}
\Ldens{a}(y) e^{-y} \leq \Ldens{a}(-y) \leq \Ldens{a}(y).
\end{align*}
We show that such a bound indeed stays preserved under MS processing at the
check nodes. The equivalent question at the variable nodes is an open question.

What are the fundamental performance limits under MS decoding? Under BP
decoding an explicit optimization of the degree distribution shows that we can
seemingly get arbitrarily close to capacity by a proper choice of the degree
distribution. Is the same behavior true under MS decoding or are the
fundamental limits which can not be surpassed? In order to address this
question we implemented an optimization tool based on EXIT charts. We found that 
the gap between the best code and Shannon limit is rather large. 

In \cite{Chen} some simple improvements are proposed to the MS decoder. For
some examples, it is demonstrated that by a simple scaling of the output at the
check nodes, the performance of the MS decoder can be brought closer to that of
the BP decoder. Using the LDPC code design procedure, we also study how close
we can get to the Shannon capacity limit by using this modified MS algorithm.

The paper is organized as follows. In Section \ref{sec:prelim}, we give
relevant definitions and briefly review the MS decoding algorithm and its
density evolution analysis.  In Section \ref{sec:stabc}, we derive a sufficient
condition for stability and also discuss some properties of density which arise
in density evolution.  In Section \ref{sec:optim}, we discuss the optimization
procedure. We then present the optimization results in Section \ref{sec:res}
and finally conclude in Section \ref{sec:conclusion}.

\section{Definitions and Preliminaries}
\label{sec:prelim}
The LDPC ensemble is specified by specifying $\lambda(x) = \sum
\lambda_ix^{i-1}$ and $\rho(x) = \sum \rho_ix^{i-1}$ which represent the degree
distribution (dd) of the bit nodes and check nodes in the edge perspective,
i.e., $\lambda_i$ $(\rho_i)$ is the fraction of edges connected to a degree $i$
bit (check) node. The design rate of an LDPC ensemble is given by $1 - \sum
\frac{\rho_i}{i}/\sum \frac{\lambda_i}{i}$.

We consider transmission over a binary-input, memoryless, and symmetric (BMS)
channel. Let $L_{ch,u}$ be the log-likelihood ratio (LLR) of bit $u$ obtained
from the channel observation corresponding to bit $u$. Let $L_{cb,u,v}^{(t)}$
and $L_{bc,u,v}^{(t)}$ be the check to bit and bit to check message at
iteration $t$ corresponding to edge $(u,v)$.  We will sometimes specifically
refer to the binary input AWGN (biAWGN) channel, $Y = (1-2X)+N$, where $X \in
\{0,1\}$ is the input bit and $N$ has a Gaussian distribution with $0$ mean and
variance $\sigma^2$. In this case $L_{ch}$ is given by $2Y/\sigma^2$ and its
distribution under the all zero code word assumption is Gaussian with mean
$2/\sigma^2$ and variance $4/\sigma^2$. Finally, we denote the Bhattacharyya
constant associated to density $\ld{a}$ by $\bh\brc{\ld{a}} =
\int_{-\infty}^\infty \ld{a}(x) e^{-\frac{x}{2}} dx$ and error probability by
$\perr\brc{\ld{a}} = \int_{-\infty}^{0^-} \ld{a}(x) dx + \frac{1}{2}
\int_{0^-}^{0^+} \ld{a}(x) dx$. 

We now discuss the message passing rules for the MS decoder.
In MS decoder the bit to check message
update is given by
\begin{equation}
    L_{bc,u,v}^{(t)} = L_{ch,u}+\sum_{v':(u,v') \in {\cal E}, v' \neq v}
    L_{cb,u,v'}^{(t-1)},
\label{eqn:bitUpdate}
\end{equation}
where ${\cal E}$ is the set of edges. The check to bit message update equation is
\begin{IEEEeqnarray}{rCl}
    L_{cb,u,v}^{(t)} & = & \frac 1 \alpha \prod_{u':(u',v) \in {\cal E}, u' \neq u}
    \mbox{sgn}(L_{bc,u',v}^{(t)}) \cdot
    \min_{u':(u',v) \in {\cal E}, u' \neq u}
    |L_{bc,u',v}^{(t)}|. \IEEEeqnarraynumspace
\label{eqn:checkminsum}
\end{IEEEeqnarray}
For the MS decoder $\alpha = 1$, but we will also consider modified MS decoders
with $\alpha > 1$. 

The asymptotic performance of LDPC codes under MS decoding can be characterized
by studying the evolution of the density of the messages with iterations (see
\cite{Richardson}). Let $\ld{a}_{ch}(l)$, $\ld{b}_{t}(l)$, and $\ld{a}_{t}(l)$
be the probability density function (pdf) of channel log-likelihood ratio, the
message from check to bit and bit to check node respectively in $t^{\text{th}}$
iteration under the all zero codeword assumption.

The density evolution equation for the bit
node (corresponding to (\ref{eqn:bitUpdate})) is given by
\begin{equation}
\ld{a}_{t}(l) = \ld{a}_{ch}(l)\conv \sum \lambda_i
(\ld{b}_{t-1}(l))^{\conv (i-1)}
\end{equation}
where $\ld{a}^{\conv i}$ denotes convolution of $\ld{a}$ with itself $i$ times.
Similarly the check node side operation on densities is denoted by $\boxtimes$.
The pdf of the message at the output of check nodes employing MS (corresponding
to (\ref{eqn:checkminsum})) has been derived in \cite{Chen},
\cite{Anastasopoulos}. It is given by

\begin{IEEEeqnarray}{rCl}
\nonumber \frac{1}{\alpha}\ld{b}_{t}\left(\frac{l}{\alpha}\right)
& \triangleq & \rho\brc{\ld{a}_{t}(l)} \nonumber 
\end{IEEEeqnarray}
\begin{IEEEeqnarray}{rCl}
&& \hspace*{-0.0cm} = \sum \rho_i \frac{i-1}{2} \left[  \left(\ld{a}_{t}(l) + \ld{a}_{t}(-l)\right) \left(\int_{|l|}^{\infty}
\left(\ld{a}_{t}(x) + \ld{a}_{t}(-x)\right) dx \right)^{i-2} \right. \nonumber \\
&& \hspace*{1.0cm}\left. + \left(\ld{a}_{t}(l) -
\ld{a}_{t}(-l)\right) \left(\int_{|l|}^{\infty} \left(\ld{a}_{t}(x) -
\ld{a}_{t}(-x)\right) dx \right)^{i-2} \right]. \nonumber
\end{IEEEeqnarray}
The density evolution process is started with $\ld{b}_{0}(l) = \delta_0(l)$ and
iterative decoding is successful if the densities eventually tend to
$\delta_\infty(l)$.

\section{Stability Condition and Some Properties of the Densities}
\label{sec:stabc}
In this section we derive the stability condition under MS decoding. The
stability condition guarantees that if the density in density evolution reaches
``close'' to error free density ($\delta_\infty(l)$) then it converges to it.
We derive the stability condition by upper bounding the evolution of the
Bhattacharyya parameter in density evolution. Note that the Bhattacharyya
parameter appears naturally in the context of BP where densities are symmetric.
In this case the Bhattacharyya parameter has a very concrete meaning: it is
equal to $-\lim_{n \to \infty} \frac{1}{n} \log\brc{\perr\brc{\ld{a}^{\conv
n}}}$, where $\ld{a}$ is a symmetric density. For general densities which are
not symmetric this is no longer true but we can always compute $\bh\brc{\ld{a}} =
\int_{-\infty}^\infty \ld{a}(x) e^{-\frac{x}{2}} dx$.  The reason we use
Bhattacharyya parameter is to have a one dimensional representation of densities
and because of its property of being multiplicative on the variable node side.

In the following lemma we give a sufficient condition for stability of
$\delta_\infty(l)$. This condition turns out to be same as the stability
condition for BP (\text{Theorem 5, \cite{Richardson2}}). 
\begin{lemma}\label{lemma:stab}
Assume we are given a degree distribution pair $(\lambda, \rho)$ and that
transmission takes place over a BMS channel characterized by its $L$-density
$\Ldens{a}_{\tiny ch}$. Define $\Ldens{a}_0 = \Ldens{a}_{\tiny ch}$, and for $t
\geq 1$, define $\Ldens{a}_{t} \doteq \Ldens{a}_{\tiny ch} \conv
\lambda\brc{\rho\brc{\Ldens{a}_{t-1}}} = \Ldens{a}_{\tiny ch} \conv \sum_j
\lambda_j \brc{\sum_k \rho_k \brc{\Ldens{a_{t-1}}}^{\boxtimes
(k-1)}}^{\circledast(j-1)}$. If 
\begin{equation}\label{eqn:stabCond}
\bh\brc{\Ldens{a}_{\tiny ch}} \lambda'(0) \rho'(1) < 1, 
\end{equation}
then there exists a strictly positive constant $\xi = \xi\brc{\lambda, \rho,
\Ldens{a}_{\tiny ch}}$ such that if, for some $t \in \mathbb{N}$,
$\bh\brc{\Ldens{a}_{t}} \leq \xi$, then $\bh\brc{\Ldens{a}_{t+n}}$ as well as
$\perr\brc{\Ldens{a}_{t+n}}$ converge to zero as $n$ tends to infinity.
Conversely, if $\bh\brc{\Ldens{a}_{\tiny ch}} \lambda'(0) \rho'(1) > 1$ then
$\liminf_{t \to \infty}\perr\brc{\Ldens{a}_{t}} > 0$ with $\Ldens{a}_{0} =
\Ldens{a}_{ch}$. 
\end{lemma}
\begin{proof} By Lemma \ref{lemma:upbound} in Appendix we know that $\bh\brc{\Ldens{a}_{t}^{\boxtimes (k-1)}}
\leq (k-1) \bh\brc{\Ldens{a}_{t}}$. Thus 
\begin{IEEEeqnarray*}{rCl}
\bh\brc{\Ldens{a}_{t+1}} & = & \bh\brc{\Ldens{a}_{\tiny ch}} \lambda\brc{\bh\brc{\sum_k \rho_k \brc{\Ldens{a_{t}}}^{\boxtimes
(k-1)}}},\\ 
 & \leq &  \bh\brc{\Ldens{a}_{\tiny ch}} \lambda\brc{\rho'(1) \bh\brc{\Ldens{a}_{t}}}.
\end{IEEEeqnarray*}
Expanding the last equation around zero, we get 
\begin{IEEEeqnarray*}{rCl}
& = & \bh\brc{\Ldens{a}_{\tiny ch}} \lambda'(0) \rho'(1) \bh\brc{\Ldens{a}_{t}} + O\brc{\bh\brc{\Ldens{a}_{t}}^2}.
\end{IEEEeqnarray*}
Since $\bh\brc{\Ldens{a}_{\tiny ch}} \lambda'(0) \rho'(1)$ is assumed to be a
constant less than 1, we can choose a sufficiently small $\xi=\xi\brc{\lambda,
\rho, \Ldens{a}_{\tiny ch}}$ such that if $\bh\brc{\Ldens{a}_{t}} \leq \xi$,
then $\bh\brc{\Ldens{a}_{\tiny ch}} \lambda'(0) \rho'(1) + O\brc{\Ldens{a}_{t}}
\leq \epsilon < 1$. Therefore if for some $t \in \mathbb{N}$,
$\bh\brc{\Ldens{a}_{t}} \leq \xi$, then $\bh\brc{\Ldens{a}_{t+n}} \leq
\epsilon^n \bh\brc{\Ldens{a}_{t}}$, which converges to zero as $n$ tends to
infinity. As 
\begin{multline*}
\perr\brc{\ld{a}_{t+n}}  =    \int_{-\infty}^{0^-} \ld{a}(x) dx + \frac{1}{2} \int_{0^-}^{0^+} \ld{a}(x) dx \\
			 \leq \int_{-\infty}^{0^+} \ld{a}(x) e^{-\frac{x}{2}} dx \leq \bh\brc{\ld{a}_{t+n}}, 
\end{multline*}
so $\perr\brc{\ld{a}_{t+n}}$ also converges to zero. 

For the converse statement, the stability condition in Eqn(\ref{eqn:stabCond})
is a necessary condition for BP decoding to be successful. Hence by the
optimality of BP decoding on a tree it is also a necessary condition for MS
decoding to be successful.  
\end{proof}

In proving the sufficiency of the stability condition we used the Bhattacharyya
parameter as the functional to project densities to one dimension.  However we
could have used any other functional of the form $\bh_{\alpha}\brc{\ld{a}} =
\mathbb{E}\left[ e^{-\alpha X}\right], \alpha > 0$ which is multiplicative on
the variable node side. Lemma \ref{lemma:upbound} stays valid for any such
functional. Therefore, we get a general stability condition that reads
$\bh_{\alpha}\brc{\ld{a}_{ch}} \lambda'(0) \rho'(1) < 1$. However, as
$\ld{a}_{ch}(x)$ is a symmetric density, $\bh_{\alpha}\brc{\ld{a}_{ch}} \geq
\bh\brc{\ld{a}_{ch}}$. This implies that the sufficient condition for $\alpha
\neq \frac{1}{2}$ is weaker than the condition corresponding to Bhattacharyya
parameter. 

Note that the converse in Lemma \ref{lemma:stab} is partial. It does not say
that the condition in Eqn(\ref{eqn:stabCond}) is necessary for the density to
converge to $\delta_\infty(l)$ if for some $t$ the density $\ld{a}_t$ is
``close'' to $\delta_\infty(l)$. However the following observation suggests
that this indeed should be the necessary condition. Suppose we evolve the
density $2 \epsilon \delta_0(l) + (1-2 \epsilon) \delta_\infty(l)$ under the MS
decoder. Then it again follows by the arguments of Theorem 5 in
\cite{Richardson2}) that for the density to converge to $\delta_\infty(l)$ the
necessary condition is $\bh\brc{\ld{a}_{\tiny ch}} \lambda'(0) \rho'(1) < 1$.
For the BP decoder we know that $2 \epsilon \delta_0(l) + (1-2 \epsilon) \delta_\infty(l)$ 
is the ``best'' density (in the sense of degradation) with error probability $\epsilon$. 
However for the MS decoder this is not the case. Hence we can not conclude that 
Eqn(\ref{eqn:stabCond}) is a necessary condition. 

The BP densities satisfy the symmetry condition $\ld{a}(x) = \ld{a}(-x)
e^{x}$.  The densities which arise in MS decoder do not satisfy the
symmetry property. However, we have observed empirically that the densities
satisfy the property that $\ld{a}(x) \geq \ld{a}(-x)$ and  $\ld{a}(x) \leq
\ld{a}(-x) e^{x}$, $x > 0$. In the following lemma we
prove that these properties remain preserved on the check node side.
\begin{lemma}\label{lemma:denschar}
Let $\ld{a}(x)$ and $\ld{b}(x)$ be two densities which satisfy the property that 
$\ld{a}(x) \geq \ld{a}(-x), \ld{b}(x) \geq \ld{b}(-x)$ and 
$\ld{a}(x) \leq e^x \ld{a}(-x), \ld{b}(x) \leq e^{x} \ld{b}(-x)$ for $\forall x > 0$. 
Let $\ld{c}(x) = (\ld{a} \boxtimes \ld{b}) (x)$. Then $\ld{c}(x) \geq \ld{c}(-x)$ and 
$\ld{c}(x) \leq e^x \ld{c}(-x)$.
\end{lemma}
\begin{proof}
Let $A$ and $B$ be random variables having density $\ld{a}$ and $\ld{b}$ respectively. Then 
\begin{IEEEeqnarray*}{rCl}
\ld{c}(x) & = & \ld{a}(x) \prob\brc{B > |x|} + \ld{b}(x) \prob\brc{A > |x|} + \\
& & \ld{a}(-x) \prob\brc{B < -|x|} + \ld{b}(-x) \prob\brc{A < -|x|}.
\end{IEEEeqnarray*}
Thus 
\begin{IEEEeqnarray*}{rCl}
\ld{c}(x)-\ld{c}(-x) & = & \brc{\ld{a}(x)-\ld{a}(-x)} \brc{\prob\brc{B > x}-\prob\brc{B < -x}} + \\
& & \brc{\ld{b}(x)-\ld{b}(-x)} \brc{\prob\brc{A > x}-\prob\brc{A < -x}},\\
& \geq & 0.
\end{IEEEeqnarray*}
Similarly,
\[
\ld{c}(-x)-e^{-x} \ld{c}(x) = \brc{\ld{a}(-x)-e^{-x} \ld{a}(x)} \prob\brc{B > x} + 
\]
\[
 \brc{\ld{b}(-x)-e^{-x} \ld{b}(x)} \prob\brc{A > x} + \brc{\ld{a}(x)-e^{-x} \ld{a}(-x)} \prob\brc{B < -x} 
\]
\[
+ \brc{\ld{b}(x)-e^{-x} \ld{b}(-x)} \prob\brc{A < -x},
\]
which is greater than or equal to zero by the assumption.
\end{proof}
Proving Lemma \ref{lemma:denschar} for the variable node side is still an open question.
\section{Optimization Procedure}
\label{sec:optim}
\subsection{ EXIT Charts}
EXIT charts \cite{tenBrink} were proposed as a low complexity
alternative to design and analyze LDPC codes. Typically by assuming that the
density of the messages exchanged during iterative decoding is Gaussian, the
problem of code design can be reduced to a curve fitting problem which can be
done using linear programming. If the Gaussian assumption is exact, this
technique is shown to be optimal in \cite{Bhattad}.  In \cite{Amraoui}, a fast
procedure is proposed that uses a combination of EXIT charts and density
evolution to design LDPC codes. The basic idea is to perform the design in
steps, where, in each step, the LDPC code ensemble is optimized using EXIT
charts using the densities of the messages obtained from density evolution of
the ensemble obtained in the previous step. In this paper, we use a similar
idea to design LDPC codes for MS decoding.

\begin{figure}
\centering
\includegraphics[width = 3.5in]{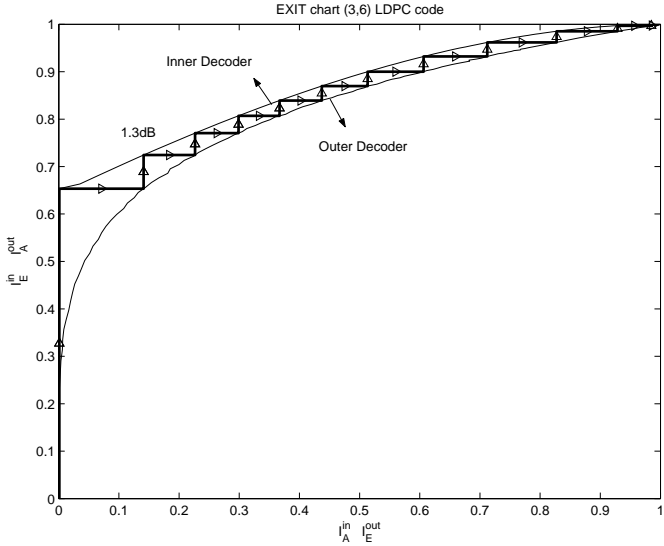}
\caption{EXIT curves of the two component codes corresponding to the
(3,6) LDPC code transmitted over an AWGN channel with $E_b/N_0 =
1.3$ dB} \label{fig:EXIT_chart_example}
\end{figure}

An EXIT curve of a component decoder is a plot of the mutual information
corresponding to the extrinsic output expressed as a function of the
mutual information corresponding to the a priori input (message coming from
the other component decoder).
Usually, it is assumed that the a priori information is from an
AWGN channel of signal-to-noise ratio $1/\sigma^2$ and the EXIT
curve is obtained by calculating the
input and output mutual information for  $\sigma^2$ varying from $0$
to $\infty$. In an EXIT chart, the EXIT curves of one component code
and the flipped EXIT curve of the other component code are plotted.
Using this chart, we can predict the path taken by the iterative decoder
as shown in Fig. \ref{fig:EXIT_chart_example}. It has been observed that the
actual path taken and the path predicted from EXIT charts are quite close.
Based on this observation, LDPC codes can be designed as follows.

Let $I_b(I_A,i)$ ($I_c(I_A,i)$) be mutual information corresponding
to the extrinsic output of bit (check) node of degree $i$ when the
a priori mutual information is $I_A$. The mutual
information $I$ can be calculated from the conditional distribution
$f(l)$ using
\begin{equation}
    I = \int_{-\infty}^{\infty} f(l)\log_2
\frac{2f(l)}{f(l)+f(-l)}\ dl.
\end{equation}

The EXIT curve of the bit nodes and the check nodes
is given by $I_b = \sum \lambda_i
I_b(I_A,i)$ and $I_c = \sum \rho_i I_c(I_A,i)$ respectively.
Usually both $I_b$ and $I_c$ are increasing function of $I_A$. The
convergence condition, based on the assumption on the message
density, states that the EXIT curve of the bit nodes should lie above that of the check nodes
for the iterative decoder to converge to the correct codeword, i.e., $I_b(I_A)
> I_c^{-1}(I_A)$ or equivalently $I_b^{-1}(I_A) < I_c(I_A)$ for all
$I_A$ where $I_c(I_c^{-1}(I_A))=I_A$. For a fixed $\rho(x)$, the
problem of code design can then be stated as the following linear program
\begin{align}
    &\max& \sum \lambda_i/i \nonumber\\
    &\mbox{subject to:}& \sum \lambda_i = 1 , \lambda_i \geq 0, \nonumber\\
  && \sum \lambda_i I_b(I_A,i) > I_c^{-1}(I_A)\ \  \forall I_A \in
[0,1).
\end{align}
Note that maximizing the objective function corresponds to maximizing
the rate. A
similar linear program can be written for optimizing $\rho(x)$ for a
given $\lambda(x)$.

\subsection{Fixed Channel} 
We consider the problem of finding LDPC codes for a given BMS channel such that
reliable communication is possible with the MS decoding algorithm. We are
interested here in the performance when the block length goes to infinity.  Our
goal is to maximize the rate of transmission. Towards achieving this goal we
first pick an LDPC code such that it converges to error free density for the specified channel. 

Starting from the initial ensemble, the LDPC code ensemble is optimized in
several steps. In each step, the basic idea is to design the codes using EXIT
charts. However, instead of using the Gaussian assumption on the input
densities, the input density in a particular step of the optimization process
is assumed to be the same as the density obtained by using the density
evolution procedure for the ensemble obtained in the previous step. The
inherent assumption is that the input densities do not change much in one step
of the optimization procedure and therefore the approximate EXIT curves
obtained using the previous densities are close to the actual EXIT curves. Note
that this assumption is different from the assumption that the density at
iteration $i$ for a  particular optimization step is same as the density at
iteration $i$ in the next optimization step.  If we denote the densities at
iteration $i$ by $\ld{a}_i$ and consider a family of densities that includes
$\{ \gamma \ld{a}_i + (1-\gamma) \ld{a}_{i+1}, \gamma \in [0,1]\}$ then the
assumption made is that the family of densities does not change much in one
step of the optimization. We could sample many points in this family to enforce
the condition in the linear program that the EXIT curves do not intersect.
However, we sample only at points $\ld{a}_i$. This is usually sufficient since
if the old EXIT curves are close to each other, then we get many samples there
and at other points we have more leeway so we can sample fewer times.

In each step of the optimization procedure, we generate a new dd pair from the previous dd pair in
two sub-steps. In the first sub-step we change $\lambda(x)$ keeping
$\rho(x)$ constant and in the next sub-step we change $\rho(x)$
while keeping $\lambda(x)$ the same. The first sub-step is as
follows. We choose $\rho(x) = \rho_{old}(x)$ and optimize
$\lambda(x)$ as follows. We perform density evolution with the dd
pair $(\lambda_{old},\rho_{old})$ and at the end of each iteration
store $I_b^l(d)$ which is the mutual information corresponding to
the extrinsic output of a bit node of degree $d$ at the end of
iteration $l$. The optimization then reduces to the following linear
program.
\begin{IEEEeqnarray}{rCl}
  &&\qquad\max \sum \lambda_i/i \nonumber\\
  &&\qquad\sum \lambda_i = 1, \lambda_i \geq 0, \nonumber\\ 
  &&\qquad\sum \frac{\lambda_i}{i} \geq \sum \frac{\rho_i}{i}, \nonumber \\
  &&\qquad\sum \lambda_i I_b^l(i) > \sum \lambda_{old,i} I_b^{l-1}(i) \nonumber \\
  &&\qquad\ \ + \beta \sum \lambda_{old,i} (I_b^{l}(i)-I_b^{l-1}(i))\ \
     \beta \in [0,1) \ \forall l, \ \ \ \
  \label{eqn:exitmatching}
 \\&&\qquad -\delta \leq \lambda_i - \lambda_{old,i} \leq \delta \qquad
  \forall \ i,  \label{eqn:maxchange}\\
  && \qquad\lambda_2 \leq \frac{1}{\bh\brc{\ld{a}_{ch}} \rho'(1)} \ \ \  \  \mbox{(from
(\ref{eqn:stabCond}))}.
  \label{eqn:stabCond2}
\end{IEEEeqnarray}
Before we explain the constraints, we note that the cost function
corresponds to maximizing the rate and that the old dd pair
satisfies the constraints and therefore the resulting rate is always
larger than the old rate.

The Constraint (\ref{eqn:exitmatching}) basically represents the
condition that the EXIT curve corresponding to the bit nodes should
lie above that of the check nodes. The quantity $\sum
\lambda_{old,i} (I_b^{l}(i)-I_b^{l-1}(i))$ is the gap between the
two EXIT curves corresponding to the old dd pair. The constant
$\beta$ determines how much change in the gap is allowed. If $\beta$
is chosen to be 0 the gap between the curves can become zero while
if $\beta$ is chosen to be one the gap is kept the same.

By choosing a smaller $\beta$ we weaken the constraints and
therefore get a larger rate. However, since
the dd pair changes, the input densities also change and therefore
the actual EXIT curves change. Since the gap between the approximate
EXIT curves (one obtained using the previous densities) is smaller
with smaller $\beta$, the chances of the actual EXIT curves
intersecting increases. We choose some value of $\beta$, perform the
density evolution with the new dd pair and check if it converges. If
it does, we accept the new ensemble and go to the second sub-step. If it
does not converge, we increase $\beta$ and repeat this sub-step.

The Constraint (\ref{eqn:maxchange}) is introduced so that the
degree distributions do not change much in an iteration which in turn will
ensure that the input densities and the resulting EXIT curves do not
change significantly.

The Constraint (\ref{eqn:stabCond2}) is the stability condition. 
For the modified MS algorithm with $\alpha > 1$, we replace the stability 
condition by the condition $\lambda_2\rho'(1) < 1$. 

In the second sub-step we perform the density evolution with the dd
pair obtained in the previous sub-step and store $I_c^l(d)$ which is
the mutual information corresponding to the extrinsic output of a
degree $d$ check node at the end of iteration $l$. A linear program,
similar to that discussed before, can then be used to optimize the rate.
As mentioned before, the rate keeps increasing with each step of the
optimization process. We stop the optimization when the increase in
rate becomes insignificant.

The linear program discussed above can be easily modified for the case when 
we have a fixed rate and we want to find a code with better threshold.  
This optimization procedure is available on-line at \cite{Bhattad2}.

\section{Optimization Results}
\label{sec:res}
We used the optimization procedure discussed in this paper to design
LDPC codes for MS. For fixed rate optimization scheme the gap to capacity
varied significantly depending on the average right degree chosen.
For the fixed channel optimization procedure, the final gap to capacity
depended on the initial profile with which the optimization procedure was started however
the variations were observed to be lesser than that in fixed rate optimization.

\begin{figure}
\centering
\includegraphics[width = 3.5in]{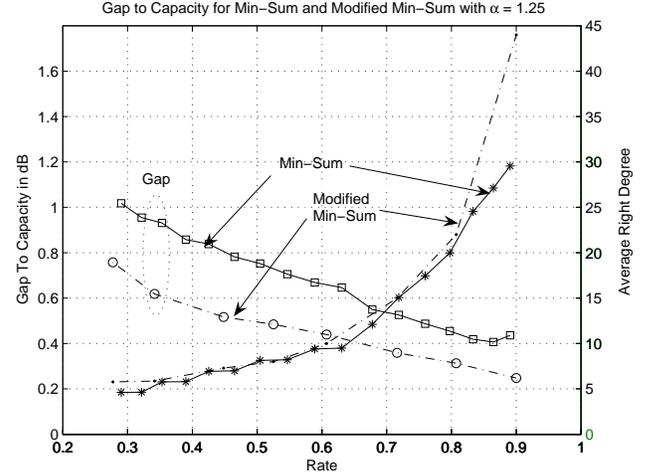}
\caption{ Gap to capacity of some optimized profiles}
\label{fig:gaptocap}
\end{figure}

In Fig. \ref{fig:gaptocap} we show the gap to capacity and the average right degree
corresponding to LDPC codes optimized for MS decoding and modified MS decoding with
$\alpha = 1.25$. The fixed channel optimization
procedure was used to obtain these points. We observe that the gap decreases as the rate
increases but it is still quite far from the Shannon capacity limit.

Comparison of the threshold of LDPC codes designed for BP but used with
MS and the threshold of codes designed for MS shows
that significant gains are obtained by using codes specifically designed
for MS. For example, the best rate 0.5 code designed for BP
from \cite{Amraoui2} has a threshold of 1.91 dB with MS which is
0.97 dB worse than the best threshold we obtained for LDPC codes that were
optimized for MS \cite{Bhattad2}.

\section{Conclusion}
\label{sec:conclusion}
We derived a sufficient condition for the stability of the fixed point
$\delta_\infty(l)$ which is also a necessary condition for the density
evolution to converge to $\delta_\infty(l)$ when initiated with channel
log-likelihood ratio density.  It remains an open question whether this condition is
also necessary for the stability of fixed point $\delta_\infty(l)$ subjected to 
 local perturbation.

We have discussed some properties of densities which are observed to be
empirically true. We proved that these properties remain preserved on the check
node side. It remains to be seen if the same thing can be proved for the
variable node side.
 
We presented a simple procedure to optimize LDPC codes for MS decoding. To the
best of our knowledge, the obtained codes are the best codes reported so far
for MS decoding and they perform significantly better than codes that were
designed for BP but are decoded using MS. However, their performance is quite
far from the capacity limit and it remains to be seen if the gap is due to the
sub-optimality of the design procedure. On the other hand if the gap is due to the 
inherent sub-optimality of MS, it will be an interesting research direction to explain the 
gap by information theoretic reasoning. 

\appendix
\begin{lemma}\label{lemma:upbound}
Let $\Ldens{a}$ and $\Ldens{b}$ be two densities and $\Ldens{c} = \Ldens{a} \boxtimes \Ldens{b}$. Then 
\[
	\bh\brc{\Ldens{c}} \leq \bh\brc{\Ldens{a}} + \bh\brc{\Ldens{b}}.
\] 
\end{lemma}
\begin{proof}
For the sake of simplicity, in the proof we assume that densities $\ld{a}$ and
$\ld{b}$ are absolutely continuous. However the proof also works in the general case. 
Let $X$ and $Y$ be two random variables
with densities $\Ldens{a}$ and $\Ldens{b}$ respectively and $Z =
\mathrm{sign}\brc{X} \mathrm{sign}\brc{Y} \min\brc{|X|, |Y|}$. Then
$\bh\brc{\Ldens{c}} = \mathbb{E}\sqbrc{e^{-\frac{Z}{2}}}$,

\begin{IEEEeqnarray}{rCl}
\bh\brc{\Ldens{c}} & = & \int_{-\infty}^\infty \int_{-\infty}^\infty e^{-\frac{\mathrm{sign}\brc{x} 
\mathrm{sign}\brc{y} \min\brc{|x|, |y|}}{2}} \Ldens{a}(x) \Ldens{b}(y) dy dx \nonumber, \\
& = & \int_{0}^\infty \int_{0}^\infty \brc{\Ldens{a}(x) \Ldens{b}(y) + \Ldens{a}(-x) \Ldens{b}(-y)}e^{-\frac{\min(x, y)}{2}} dy dx + \nonumber \\
& &  \int_{0}^\infty \int_{0}^\infty \brc{\Ldens{a}(x) \Ldens{b}(-y) + \Ldens{a}(-x) \Ldens{b}(y)}e^{\frac{\min(x, y)}{2}} dy dx, \nonumber \\ 
& \stackrel{(a)}{=} & \int_{0}^\infty \int_{0}^x \frac{g(x,y)}{g(x,y)} \left\{ \brc{\Ldens{a}(x) \Ldens{b}(y) + \Ldens{a}(-x) \Ldens{b}(-y)}e^{-\frac{y}{2}} \right.\nonumber \\ 
& & \left. + \brc{\Ldens{a}(x) \Ldens{b}(-y) + \Ldens{a}(-x) \Ldens{b}(y)} e^{\frac{y}{2}}\right\} dy dx + \nonumber  
\end{IEEEeqnarray}
\begin{IEEEeqnarray}{rCl}
& & \int_{0}^\infty \int_{x}^\infty \frac{g(x,y)}{g(x,y)} \left\{ \brc{\Ldens{a}(x) \Ldens{b}(y) + \Ldens{a}(-x) \Ldens{b}(-y)}e^{-\frac{x}{2}} \right.\nonumber \\ 
& & \left. + \brc{\Ldens{a}(x) \Ldens{b}(-y) + \Ldens{a}(-x) \Ldens{b}(y)} e^{\frac{x}{2}}\right\} dy dx + \nonumber \\ 
& = & I_1 + I_2. \label{eqn:bhc}
\end{IEEEeqnarray}

In $(a)$ we multiply and divide by $g(x,y)=(\Ldens{a}(x) + \Ldens{a}(-x)) (\Ldens{b}(y)
+ \Ldens{b}(-y))$. Note that all the densities which arise in density evolution 
satisfy the property that $\Ldens{a}(x) = 0$ if and only if $\Ldens{a}(-x)$ is zero.
 Thus if $\Ldens{a}(x)$ or $\Ldens{b}(y)$ are equal
to zero then the integrand itself is zero and those values of $x$  and $y$ do not
contribute to the integral. Hence without lose of generality we can assume that
$a(x)$ and $b(y)$ are not zero. Now, 
\begin{IEEEeqnarray}{rCl}
\bh\brc{\Ldens{a}} & = & \int_0^\infty \brc{\Ldens{a}(x) e^{-\frac{x}{2}} + \Ldens{a}(-x) e^{\frac{x}{2}}} dx, \nonumber \\
 & \stackrel{(a)}{=} & \int_0^\infty \int_0^\infty (\Ldens{b}(y) + \Ldens{b}(-y)) \brc{\Ldens{a}(x) e^{-\frac{x}{2}} + \Ldens{a}(-x) e^{\frac{x}{2}}} dy dx, \nonumber \\
 & \stackrel{(b)}{=} & \int_0^\infty \int_0^x \frac{(\Ldens{a}(x)+\Ldens{a}(-x))}{(\Ldens{a}(x)+\Ldens{a}(-x))} (\Ldens{b}(y) + \Ldens{b}(-y)) \nonumber\\ 
& & \brc{\Ldens{a}(x) e^{-\frac{x}{2}} + \Ldens{a}(-x) e^{\frac{x}{2}}} dy dx + \nonumber \\
 &  & \int_0^\infty \int_x^\infty \frac{(\Ldens{a}(x)+\Ldens{a}(-x))}{(\Ldens{a}(x)+\Ldens{a}(-x))} (\Ldens{b}(y) + \Ldens{b}(-y)) \nonumber \\ 
& & \brc{\Ldens{a}(x) e^{-\frac{x}{2}} + \Ldens{a}(-x) e^{\frac{x}{2}}} dy dx. \nonumber \\
& = & I_{\Ldens{a}1} + I_{\Ldens{a}2}. \label{eqn:bha}
\end{IEEEeqnarray}
In $(a)$ we used the fact that $\int_0^\infty \brc{b(y) + b(-y)} dy = 1$ and in
$(b)$ we multiply and divide by $(a(x) + a(-x))$.  Similarly,

\begin{IEEEeqnarray}{rCl}
\bh\brc{\Ldens{b}} & = & \int_0^\infty \int_0^x \frac{\brc{\Ldens{b}(y) + \Ldens{b}(-y)}}{\brc{\Ldens{b}(y) + \Ldens{b}(-y)}} (\Ldens{a}(x) + 
\Ldens{a}(-x)) \nonumber \\
 & & \brc{\Ldens{b}(y) e^{-\frac{y}{2}} + \Ldens{b}(-y) e^{\frac{y}{2}}} dy dx + \nonumber \\
 & & \int_0^\infty \int_x^\infty \frac{\brc{\Ldens{b}(y) + \Ldens{b}(-y)}}{\brc{\Ldens{b}(y) + \Ldens{b}(-y)}} (\Ldens{a}(x) + \Ldens{a}(-x)) \nonumber \\ 
 & & \brc{\Ldens{b}(y) e^{-\frac{y}{2}} + \Ldens{b}(-y) e^{\frac{y}{2}}} dy dx. \nonumber \\
& = & I_{\Ldens{b}1} + I_{\Ldens{b}2}. \label{eqn:bhb}
\end{IEEEeqnarray}
Note that by Eqn(\ref{eqn:bhc}, \ref{eqn:bha}, \ref{eqn:bhb}), $\bh\brc{\Ldens{c}} - \bh\brc{\Ldens{a}} - \bh\brc{\Ldens{b}} = I_1 - I_{\Ldens{a}1} - I_{\Ldens{b}1} 
+ I_2 - I_{\Ldens{a}2} - I_{\Ldens{b}2}$. We first consider $I_1 - I_{\Ldens{a}1} - I_{\Ldens{b}1}$. 
We prove that the integrand of $I_1 - I_{\Ldens{a}1} - I_{\Ldens{b}1}$ is pointwise non positive. 
As $(\Ldens{a}(x) + \Ldens{a}(-x)) (\Ldens{b}(x) + \Ldens{b}(-x))$ is a common non negative factor in the 
integrands of  $I_1, I_{\Ldens{a}1}$ and $I_{\Ldens{b}1}$, we
will not consider it. Then the remaining integrand  of $I_1 -  I_{\Ldens{a}1} - I_{\Ldens{b}1}$ is:
\begin{multline}\label{eqn:integrand}
\frac{\Ldens{a}(x) \Ldens{b}(y) e^{-\frac{y}{2}} 
+ \Ldens{a}(x) \Ldens{b}(-y) e^{\frac{y}{2}}+ \Ldens{a}(-x) \Ldens{b}(y) e^{\frac{y}{2}}+ 
\Ldens{a}(-x) \Ldens{b}(-y) e^{-\frac{y}{2}}}{(\Ldens{a}(x)+\Ldens{a}(-x)) (\Ldens{b}(y)+\Ldens{b}(-y))}  \\
- \frac{\Ldens{a}(x) e^{-\frac{x}{2}} + \Ldens{a}(-x) e^{\frac{x}{2}}}{\Ldens{a}(x) + \Ldens{a}(-x)} - 
\frac{\Ldens{b}(y) e^{-\frac{y}{2}} + \Ldens{b}(-y) e^{\frac{y}{2}}}{\Ldens{b}(y) + \Ldens{b}(-y)}.
\end{multline}
Define $q=\frac{\Ldens{b}(-y)}{\Ldens{b}(y)+\Ldens{b}(-y)}$,
$p=\frac{\Ldens{a}(-x)}{\Ldens{a}(x)+\Ldens{a}(-x)}$. Now we can write Eqn (\ref{eqn:integrand}) as 
\begin{align}
((1-p) (1-q) + p q) e^{-\frac{y}{2}} + (p (1-q) + q (1-p)) e^{\frac{y}{2}} & & \nonumber\\
 -(1-q) e^{-\frac{y}{2}} - q e^{\frac{y}{2}} - (1-p) e^{-\frac{x}{2}} - p e^{\frac{x}{2}} & & \nonumber 
\end{align}
\begin{align}
=  p (1-2 q) \brc{e^{\frac{y}{2}} - e^{-\frac{y}{2}}} - p e^{\frac{x}{2}} - (1-p) e^{-\frac{x}{2}}. & & \label{eqn:integrandpq}
\end{align}
The Eqn(\ref{eqn:integrandpq}) is exactly the Eqn(\ref{eqn:bhgbsc}) in Lemma
\ref{lemma:gbsc} which is proved to be non positive. Also note that as required
by Lemma \ref{lemma:gbsc}, $y \leq x$ and $y$ is associated with $q$. 
The integrand of $I_2 - I_{\Ldens{a}2} - I_{\Ldens{b}2}$ can also be reduced to 
Eqn(\ref{eqn:bhgbsc}) in Lemma \ref{lemma:gbsc}.  Hence we prove that
$\bh\brc{\Ldens{c}} \leq \bh\brc{\Ldens{a}} + \bh\brc{\Ldens{b}}$.
\end{proof}

We define a Generalized BSC density by, 
\[
\Ldens{a}_{\tiny \gbsc(p, x)}(z) = p \delta_{-x}(z) + (1-p) \delta_x(z).
\]
\begin{lemma}\label{lemma:gbsc}
Consider $\Ldens{a}_{\gbsc(p, x)}(z)$, $\Ldens{a}_{\gbsc(q, y)}(z)$ and 
$\Ldens{c}(z)=\Ldens{a}_{\gbsc(p, x)}(z) \boxtimes \Ldens{a}_{\gbsc(q, y)}(z)$. Then 
\[
	\bh\brc{\Ldens{c}} - \bh\brc{\Ldens{a}_{\gbsc(p, x)}} - \bh\brc{\Ldens{b}_{\gbsc(q, y)}} \leq 0.
\]
\end{lemma}

\begin{proof}
With out loss of generality we can assume that $y \leq x$. Then 
\[
\Ldens{c}(z) = (p (1-q) + q (1-p)) \delta_{-y}(z) + (p q + (1-p) (1-q)) \delta_{y}(z).
\]
Now,
\begin{align}
\bh\brc{\Ldens{c}} - \bh\brc{\Ldens{a}_{\gbsc(p, x)}} 
- \bh\brc{\Ldens{b}_{\gbsc(q, y)}}  =  & & \nonumber \\
(p (1-q) + q (1-p)) e^{\frac{y}{2}} + (p q + (1-p) (1-q)) e^{-\frac{y}{2}} & & \nonumber \\
- p e^{\frac{x}{2}} - (1-p) e^{-\frac{x}{2}} - q e^{\frac{y}{2}} - (1-q) e^{-\frac{y}{2}}. & & \nonumber \\
 =  p (1-2 q) \brc{e^{\frac{y}{2}} - e^{-\frac{y}{2}}} - p e^{\frac{x}{2}} - (1-p) e^{-\frac{x}{2}}, & & \nonumber \\ 
\leq 0, \hspace*{5.6cm} \quad && \label{eqn:bhgbsc}
\end{align}
because $1 -2 q \leq 1$ and $y \leq x$, we have $ p (1-2 q) \brc{e^{\frac{y}{2}} - e^{-\frac{y}{2}}} - p e^{\frac{x}{2}} \leq 0$. Thus we have prove the desired statement. 
\end{proof}

\end{document}